\documentclass[aps,pre,twocolumn,showpacs,superscriptaddress]{revtex4}
\usepackage{epsfig,xcolor,graphics}
\usepackage{amsmath, mathrsfs}
\usepackage{amssymb}
\usepackage{siunitx}

\definecolor{cream}{RGB}{222,217,201}

\definecolor{myc}{rgb}{1,0,0}  

\graphicspath{{FIGURES/}}

\begin{document}
\title{Spinning elastic beads: a route for simultaneous measurements of shear modulus and interfacial energy of soft materials}
\author{Alessandro Carbonaro}
\thanks{These three authors contributed equally}
\affiliation{Laboratoire Charles Coulomb, Universit\'e de Montpellier and CNRS, France}
\author{Kennedy-Nexon Chagua-Encarnacion}
\thanks{These three authors contributed equally}
\affiliation{Laboratoire Charles Coulomb, Universit\'e de Montpellier and CNRS, France}
\author{Carole-Ann Charles}
\thanks{These three authors contributed equally}
\affiliation{Laboratoire Charles Coulomb, Universit\'e de Montpellier and CNRS, France}
\author{Ty Phou}
\affiliation{Laboratoire Charles Coulomb, Universit\'e de Montpellier and CNRS, France}
\email{ty.phou@umontpellier.fr}
\author{Christian Ligoure}
\email[Corresponding author: ]{christian.ligoure@umontpellier.fr}
\affiliation{Laboratoire Charles Coulomb, Universit\'e de Montpellier and CNRS, France}
\author{Serge Mora}
\email[Corresponding author: ]{serge.mora@umontpellier.fr}
\affiliation{Laboratoire de M\'ecanique et G\'enie Civil, Universit\'e de Montpellier and CNRS, France}
\author{Domenico Truzzolillo}
\email[Corresponding author: ]{domenico.truzzolillo@umontpellier.fr}
\affiliation{Laboratoire Charles Coulomb, Universit\'e de Montpellier and CNRS, France}
\date{\today}
\begin{abstract}
Large deformations of soft elastic beads spinning at high angular velocity in a denser background fluid are investigated theoretically, numerically, and experimentally using millimeter-size polyacrylamide hydrogel particles introduced in a spinning drop tensiometer. We determine the equilibrium shapes of the beads from the competition between the centrifugal force and the restoring elastic and surface forces. Considering the beads as neo-Hookean up to large deformations, we show that their elastic modulus and surface energy constant can be simultaneously deduced from their equilibrium shape. Also, our results provide further support to the scenario in which surface energy and surface tension coincide for amorphous polymer gels.
\end{abstract}
\maketitle
\section{Introduction}

When subjected to external loads, elastic bodies change their shape due to the interplay between the applied load and the restoring forces of the material the body is made of \cite{Style2017,Bico2018}. Below the elastic limit, these are the bulk elastic forces following the material-specific stress-strain relation, and the surface forces dictated by the interfacial free energy that characterizes the interaction with the surrounding medium. Since the subtle balance between these forces stays relevant even beyond the elastic limit and determines, together with the onset of plastic events, the occurrence of material failure and permanent deformation \cite{creton_fracture_2016}, bulk and surface stresses turn out to drive the behavior of soft materials under many circumstances \cite{Style2017,Bico2018,creton_fracture_2016}. For this reason understanding the importance of these two contribution to material response is of paramount importance.

The impact of interfacial stresses on the equilibrium shape of elastic materials can be readily quantified by the elasto-capillary length $\ell$, defined as the ratio of the interfacial energy per unit area $\Gamma$ to the shear modulus $G_0$ of the body under consideration. When $\ell$ is comparable with or larger than other characteristic lengths of the system \cite{Nicolson1955,Mora_prl2010,Style2013} interfacial stresses must be taken into account to compute stationary material shapes and to predict possibly the onset of instabilities \cite{andreotti_elastocapillary_2011,evans_elastocapillary_2013,chakrabarti_direct_2013}.
This is the case for soft elastic samples with small geometric features. For example, for a hydrogel with shear modulus $G_0\sim \SI{30}{\pascal}$ and interfacial tension $\Gamma\sim \SI{30}{\milli \newton \per \meter}$, the elasto-capillary length is $\ell = \SI{1}{\milli \meter}$. Therefore, the equilibrium shapes of millimetric and submillimetric elastic particles must be necessarily affected by the interfacial contribution to their total energy. 

Despite of this general and well-grounded consideration, many important questions concerning the interplay between bulk and interfacial stresses \cite{Arora2018,Limat2018} and the very nature of the latter in amorphous solids \cite{mondal_estimation_2015,Andreotti2016} remain unanswered. For a generic material immersed in a background medium interfacial energy is the energy required to create a unit area of new surface by a division process, whereas interfacial tension is the surface stress associated with its
deformation. For Newtonian liquids, interfacial tension and interfacial energy are two strictly
equal quantities since, when a liquid interface is deformed, the distances between the molecules at the interface are not affected by the imposed deformation as molecules can move freely from the bulk to the the liquid boundaries.
It is generally not so for a solid. Since a solid surface consists of a constant number of atoms, the work done to alter the separation distance between atoms at their surface is expected to depend on this distance itself \cite{Shuttleworth1950,muller_elastic_2004,orowan_surface_1970}.
As a result, the work required to deform a material is not necessarily the same as the thermodynamic work required to create a new surface. For crystals the problem has been solved \cite{orowan_surface_1970,savina_faceting_2003} since their surface free energy is a function of the surface area itself and hence it is expected to be different from the surface tension. However, for amorphous materials, like cross-linked elastomers, the issue remains unresolved because the molecules have local mobility allowing them, at least in principe, to show liquid-like behavior: the surface reforms in response to external stimuli \cite{vaidya_synthesis_2002,hillborg_nanoscale_2004}. While this liquid-like scenario has been recently confirmed for Polydimethylsiloxane (PDMS) elastomers \cite{mondal_estimation_2015}, other experimental works pointed out that for specific soft gels \cite{Andreotti2016,style_elastocapillarity_2017,andreotti_statics_2020} the interfacial energy does depend on the surface area, or equivalently on the imposed compressive strain parallel to the surface, and, as a consequence, interfacial free energy and interfacial stress are expected to differ \cite{Shuttleworth1950,style_elastocapillarity_2017,andreotti_statics_2020}.

Moreover, for most solid materials the accurate measure of the interfacial energy is experimentally challenging, since the intimate coupling between the contributions of interfacial and bulk energies hampers the detections of effects solely due to interfacial stresses.
For instance instability thresholds \cite{Mora_prl2010,Mora_softmatter2011} and the shapes taken by softened wedges \cite{Hui2002,Mora_prl2013,Mora_JPhys2015} or ripple deformations \cite{Jagota2014} involve the coupling between surface stress and bulk elasticity through the elasto-capillary length, making impossible to determine separately the two parameters ($G_0$ and $\Gamma$). Even if one of the two parameters, {\it e.g.} the shear modulus, were determined elsewhere, a measurement relying on a single experiment is of limited accuracy.
To solve this problem, indentation tests, standard rheometry or stretching tests, based on a gradual variation of an external load could in principle be used. However, these methods involve the presence of solid-solid contact forces \cite{Chakrabarti2016,Delavoipiere2016} that typically affect the measurement and give rise to issues like slip and edge fracture. Furthermore, in the the case of ultrasoft gels the measurement of the elastic modulus through these techniques is even more troublesome since one would be confronted with issues related to insufficient instrumental accuracy.

For these reasons, unveiling effects of surface energy in soft solids remains arduous and it has not been possible to converge to any conclusive result. This motivates investigations of phenomena that originate from a non-negligible contribution of interfacial free energy in the absence of solid-solid contacts over a wide range of strains, while, at the same time, engineering strategies to fully decouple interfacial and bulk stresses would be highly desirable.\\

In this paper, we tackle this challenging task and report on a theoretical and numerical study of the deformation of soft neo-Hookean beads when they are immersed and spun in a denser background fluid. Strikingly we found that, if the interfacial energy of the beads does not depend on their deformation, the elastic and the interfacial contributions determining the bead shape can be decoupled when the strong deformation limit is reached, namely when the ratio between the two principal axis of the deformed particles is $d_{max}/d_{min}\gtrsim 2$. To check further the reliability of our results we have investigated the deformation of soft polyacrylamide beads immersed in a denser immiscible fluid and spun in the capillary of a commercial spinning drop tensiometer (SDT).

Though an SDT is usually employed to measure low liquid-liquid interfacial free energies \cite{vonnegut_rotating_1942,bamberger_effects_1984,liu_concentration_2012}, recently it has been used also for purposes ranging from the study of the relaxation dynamics of liquid drops \cite{carbonaro_spinning_2019} and the presence of an effective interfacial tension in miscible fluids \cite{carbonaro_spinning_2019,zoltowski_evidence_2007,pojman_evidence_2006}, to the characterization of the mechanical properties of thin elastic capsules \cite{pieper_deformation_1998} and viscoelastic properties of polymer melts \cite{joseph_spinning_1992,patterson_measurement_2007}.

\begin{figure}[htbp]
\centering
  \includegraphics[width=0.45\textwidth]{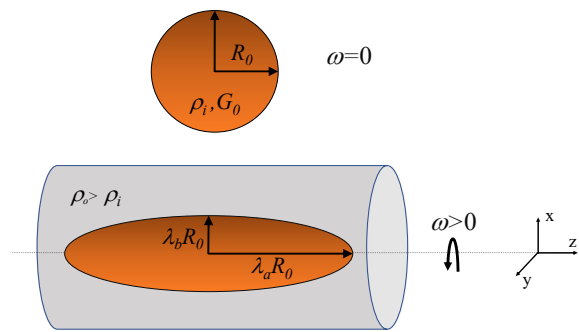}\\
  \caption{Sketch of a spherical elastic bead immersed in a liquid of higher mass density and deformed by centrifugal forcing. (a) Initial configuration of the bead at rest ($\omega=$0). (b) The elastic bead is spun solidly ($\omega>$0) with a denser fluid, both being contained in a cylindrical capillary. Centrifugal forces give rise to the reversible deformation of the bead and stabilize its position on the capillary axis.}
  \label{fig : setup}
\end{figure}

Unlike the aforementioned works, here we use an SDT to investigate the equilibrium shapes of full elastic beads with a radius of the order of one millimeter and shear modulus of the order of 10 Pa \cite{Arora2018}, for which we expect important elasto-capillary effects. When the SDT capillary is spun around its axis at a prescribed angular velocity, and once a steady state is reached, the beads spin solidly with the background fluid and the capillary itself. Since the surrounding fluid is denser than the bead, the centrifugal forces center and stretch the bead on the axis of rotation (Fig.~\ref{fig : setup}).

In this geometry, the sample bead is entirely surrounded by a liquid without any contact with other solid bodies. This is an important benefit of this geometry as the only interfacial free energy to be considered is the solid-liquid one. In addition, the external load ({\it i.e.} the centrifugal forces) can be finely tuned up to values generating large deformations ($\gtrsim500$\%) of the bead.

The remainder of this paper is organized as follows. Assuming an interfacial energy independent from the deformation, the base equations governing the equilibrium of a spinning elastic bead surrounded by a liquid spinning at the same angular velocity are derived in Sec. \ref{sec : base equations}. These equations are first solved by assuming a homogeneous (biaxial) deformation of the bead (Sec. \ref{sec : biaxial approximation}). It is shown that, within this approximation, the effects on the deformation due to the contributions of the interfacial free energy and bulk elasticity can be decoupled at high centrifugal forcing. A full resolution of the base equations is made in Sec. \ref{sec : fem} using the Finite Element method, showing the limitation of the biaxial approximation for a quantitative analysis. Interestingly, the behaviour emerging from the approximation still holds, providing a way to access to both the elastic modulus and the interfacial free energy constant of the beads. In Sec. \ref{sec : experiments} we report on experiments carried out with a commercial spinning drop tensiometer and soft polyacrylamide beads. A discussion of the main results and a comparison with the expected values for $G_0$ and $\Gamma$ follows, pointing out that soft polyacrylamide gel behavior is well capture by the assumption of a constant and strain-independent interfacial energy. Finally, in Sec. \ref{sec : conclusion} we make some concluding remarks and summarize the key results of this work.

\section{Theory}
\subsection{Equilibrium equations at finite strains} \label{sec : base equations}
The non-linear equations governing the equilibrium (steady) configuration of a rotating elastic sphere are derived considering a positive constant interfacial energy and an isotropic and incompressible neo-Hookean constitutive law. The latter is known to describe well the mechanical properties of soft polyacrylamide gels for strains up to several hundred percent \cite{Ogden1984,Suo2012,Mora2020}.

Let us consider an elastic bead of radius $R_0$, shear modulus $G_0$ and density $\rho_i$ immersed in an infinite Newtonian background fluid of density $\rho_o>\rho_i$. As the sphere is spun at angular velocity $\omega$ around one diameter (aligned along axis $z$), the bead deforms, stretching along the rotation axis to minimize its rotational energy. In the co-rotating frame the elastic force, the surface force and the centrifugal force are conservative.  The equilibrium can therefore be derived from the condition that the total potential energy is minimum. The position $\mathbf{R}$ of a material point in the deformed configuration is given as a map ${\mathbf{R}}(\mathbf{r})$ in terms of the position $\mathbf{r}$ in the undeformed configuration. For an isotropic and incompressible neo-Hookean solid, the strain energy density is:
\begin{equation}
W_{el}=\frac{G_0}{2}\mathrm{tr}\left(\mathbf{F}^T\cdot\mathbf{F}-\mathbf{1}\right),
  \label{eq:invariants}
\end{equation}
where $\mathbf{F}=\partial \mathbf{R}/\partial \mathbf{r}$ is the deformation gradient and $\mathbf{1}$ the unit
matrix. The equilibrium is governed by the minimization of the free energy
\begin{equation}
    \mathcal{E} = \Gamma\,A + \int_{\Omega_{0}} W_{el} \mathrm{d}V_{0}+ \int_{\Omega_{0}} \frac{1}{2}\Delta \rho\,\omega^2 R^2 \mathrm{d}V_{0}
     \textrm{,}
     \label{eqn : energy}
\end{equation}
where $R$ is the radial distance from the $z$-axis in the deformed configuration ($R=\mathbf{R} \cdot \mathbf{R}-\mathbf{R}\cdot \mathbf{e}_z$), $\mathrm{d}V_{0}$ is a volume element in the reference configuration, $\Omega_{0}$ is the volume occupied by the bead, $A$ is the area of the \emph{deformed} boundary and $\Delta \rho = \rho_o-\rho_i$ is the mass density contrast. 
It's worth stressing that we assume $\Gamma$ independent of the deformation and the first term on the r.h.s of Eq.~\ref{eqn : energy}, $\mathcal{E}_{\Gamma}=\Gamma A$, represents the total interfacial energy of the system. This assumption will be discussed in the light of the results reported in Sec. \ref{sec : experiments}.
The second and the third terms are respectively the elastic and the centrifugal energies \cite{Richard2018}, called later $\mathcal{E}_{\epsilon}$ and $\mathcal{E}_{\omega}$. The equilibrium is governed by the minimization of the free energy, taking into account incompressibility of the elastic material which amounts to impose that the Jacobian of the transformation is equal to one:
\begin{equation}
  \det \mathbf{F} = 1.
  \label{eqn : incompressibility}
\end{equation}

\subsection{Biaxial approximation} \label{sec : biaxial approximation}
\subsubsection{General equations within the biaxial approximation}
To a first approximation, the problem is simplified by assuming a homogeneous and biaxial deformation of the bead. In the Cartesian coordinate system (x,y,z), the applied centrifugal forcing gives rise to a prolate ellipsoid with axes $X=\lambda_b x$, $Y=\lambda_b y$ and $Z=\lambda_a z$. The stretch ratios $\lambda_a$ and $\lambda_b$ are two strictly positive constants with $\lambda_b<1$ and $\lambda_a>1$. This is sketched in Fig. \ref{fig : setup}. Eq.~\ref{eqn : incompressibility} further imposes $\lambda_a\lambda_b^2=1$.
For a neo-Hookean material, the strain energy density is $W_{el}=\frac{1}{2}G_0(\lambda_a^2+2\lambda_b^2-3)$ and the elastic energy defined in Eq.~\ref{eqn : energy} reads:
\begin{equation}
\label{eqn : energy2}
{\mathcal E}_{\epsilon}=\frac{4}{3}\pi R_0^3 \frac{1}{2}G_0(\lambda_a^2+2\lambda_b^2-3).
\end{equation}
The total interfacial energy defined in  Eq.~\ref{eqn : energy}, is:
\begin{equation}
\label{eqn : energy4}
{\mathcal E}_{\Gamma}=\Gamma A=\Gamma\left(2\pi \lambda_b^2R_0^2+ \frac{2\pi \lambda_a\lambda_b R_0^2}{e}\mathrm{arcsin}\frac{\sqrt{\lambda_a^2-\lambda_b^2}}{\lambda_a}\right).
\end{equation}
Finally the centrifugal energy, also defined in Eq.~\ref{eqn : energy} is given by:
\begin{equation}
\label{eqn : energy1}
{\mathcal E}_\omega=\frac{1}{2}\int\Delta \rho(x^2+y^2)\omega^2 dV_0 =\frac{4\pi}{15} \Delta\rho \omega^2\lambda_b^2R_0^5.
\end{equation}
Using volume conservation,  Eqs.~\ref{eqn : energy2}, \ref{eqn : energy4} and \ref{eqn : energy1} allow us writing the total reduced energy density, defined as $\varepsilon={\cal E}/(G_0 V_0)$ with $V_0$ the volume of the bead, as:
 \begin{equation}
   \begin{split}
\varepsilon= \frac{\alpha}{5}\lambda_a^{-1}+\frac{1}{2} \left(\lambda_a^2+2\lambda_a^{-1}-3\right)\\+\frac{3 \beta}{2}\left(\lambda_a^{-1}+\frac{\lambda_a^2}{\sqrt{\lambda_a^3-1}}\mathrm{arcsin}\sqrt{\frac{\lambda_a^3-1}{\lambda_a^3}}\right),
\end{split}
\label{eqn : lambdaa}
 \end{equation}
 with
\begin{equation}
  \alpha=\frac{\Delta \rho R_0^2\omega^2}{G_0}
  \label{eqn : alpha}
\end{equation}
and
\begin{equation}
  \beta=\frac{\Gamma}{G_0 R_0}
\label{eqn : beta}
\end{equation}
being two characteristic dimensionless numbers. In particular, $\alpha$ is the Cauchy number and results from the balance between inertia and elastic energy, while $\beta$ is the ratio of the elasto-capillary length to the bead radius.

\begin{figure}[htbp]
\centering
  \includegraphics[width=0.4\textwidth]{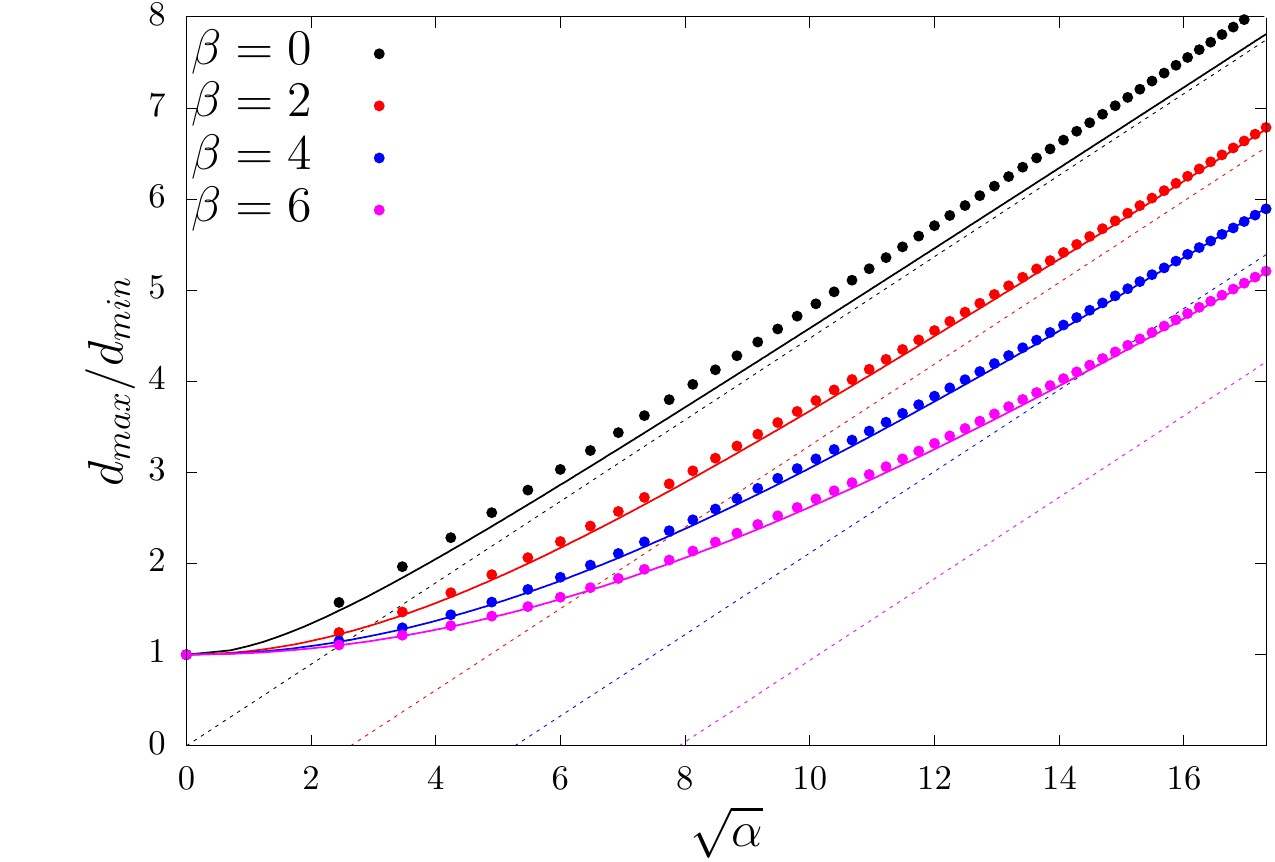}
  \caption{Length-to-width ratio $d_{max}/d_{min}$ of a deformed bead, as a function of  $\sqrt{\alpha}$ for different values of $\beta$ (color coded). Solid lines are predictions from the biaxial approximation (minimization of Eq.~\ref{eqn : lambdaa}). Asymptotes in the large $\alpha$ limit (Eq.~\ref{eqn:energy8}) are plotted with dashed lines. Filled circles are the results of FE calculations discussed in Sec.~\ref{sec : fem}.}
  \label{fig : compare}
\end{figure}

For a given set $(\alpha, \beta)$, the equilibrium shape is given by the minimization of $\varepsilon$ with respect to $\lambda_a$, {\it i.e.} by the solution of the nonlinear algebraic equation ${\frac{d\varepsilon}{d\lambda_a}=0}$, which can be obtained numerically. The deformation parameter defined as the ratio of the length ($d_{max}$) to the width ($d_{min}$) of the deformed shape, or equivalently $\lambda_a^{3/2}$,  obtained numerically by minimizing Eq. \ref{eqn : lambdaa}, is plotted in Fig. \ref{fig : compare}. Hereafter we derive the analytical expression for the deformation parameter $\lambda_a^{3/2}$ in the two limiting cases of small and large deformations and we show that when the latter are attained a simultaneous measurement of the elastic modulus and the interfacial free energy is feasible.

\subsubsection{Small deformation limit}

Let us first elucidate the behavior of the stretch ratio $\lambda_a$ in the weak deformation limit, corresponding to Cauchy numbers $\alpha \ll 1$. In this case we can safely write $\lambda_a=1+\xi$, with $0<\xi \ll 1$ and approximate the reduced energy density as follows:
\begin{equation}
\label{eqn:energy10}
\epsilon\simeq\frac{1}{5}\alpha+3\beta-\frac{1}{5}\xi\alpha+\xi^2(\frac{1}{5}\alpha+\frac{3}{2}+\frac{6}{5}\beta).
\end{equation}
Minimizing with respect to $\xi$ brings to the following equilibrium deformation with respect to the bead at rest:
\begin{equation}
\label{eqn:energy13}
\lambda_a^{3/2}-1\simeq\left[\frac{\alpha}{10+8\beta}\right].
\end{equation}
Eq.~\ref{eqn:energy13} cannot be used to determine separately $G_0$ and $\Gamma$ from a single measurement of $\lambda_a^{3/2}$ as a function of $\omega$, since the deformation cannot be expressed as the sum of two (or more) terms each containing only $\alpha$ or $\beta$ separately.

\subsubsection{Large deformation limit}
For $\alpha \gg 1$ and $\lambda_a \gg 1$, Eq.~\ref{eqn : lambdaa} can be approximated by the algebraic sum of three terms:
\begin{equation}
\label{eqn:energy6}
\epsilon\simeq \frac{1}{2}\lambda_a^2+\frac{3\pi}{4}\beta \lambda_a^{1/2}+\frac{\alpha}{5}\lambda_a^{-1}.
\end{equation}
The minimization with respect to $\lambda_a$ brings to:
\begin{equation}
\label{eqn:energy7b}
d_{max}/d_{min}=\lambda_a^{3/2}=\frac{1}{2}\left[-\frac{3}{8}\pi\beta+\sqrt{\frac{9}{64}\pi^2\beta^2+\frac{4}{5}\alpha}\right].
\end{equation}
Further expanding Eq. \ref{eqn:energy7b} for $\alpha\gg$1, we obtain:
\begin{equation}
\label{eqn:energy8}
d_{max}/d_{min}=\sqrt{\frac{\alpha}{5}}-\frac{3}{16}\pi\beta.
\end{equation}
These asymptotes are plotted together with the complete expressions of $d_{max}/d_{min}$ in Fig.~\ref{fig : compare} for different values of $\beta$. Note that due to the limited range of $\alpha$, chosen accordingly with experiments introduced in Sec. \ref{sec : experiments}, the differences between the complete expression and the asymptotes remain significant and increases for increasing values of $\beta$.
Quite interestingly Eq. \ref{eqn:energy8} decouples $\alpha$ and $\beta$, {\it i.e.} the effects of elasticity and interfacial energy on the bead deformation. In other words, for large centrifugal forcing, the deformation parameter $\lambda_a^{3/2}$ of a bead is proportional to the rotation speed with a proportionality constant equal to $R_0\sqrt{\frac{\Delta\rho}{G_0}}$. $R_0$ and $\Delta \rho$ being easily known {\it a priori}, $G_0$ can then be determined by considering the slope of $\lambda_a^{3/2}$ versus $\omega$; next, the evaluation of a (virtual) intercept equal to $-\frac{3\pi\Gamma}{16G_0 R_0}$ brings to the measurement of $\Gamma$.

$G_0$ and $\Gamma$ can then be recovered by considering the large deformation limit with the biaxial approximation. In the following, we show that the approximation is not accurate enough to get precise values of these two quantities. Notwithstanding this, the main result stays valid: it is possible to determine both $G_0$ and $\Gamma$ by considering the large deformation limit of a spinning bead.

\subsection{Resolution using the Finite Element method} \label{sec : fem}
This section is devoted to the minimization of Eq.~\ref{eqn : energy} with the incompressibility condition  (Eq.~\ref{eqn : incompressibility}), using the Finite Element (FE) method. \\

We seek the displacement $\mathbf{u}=\mathbf{R}-\mathbf{r}$ by minimizing the augmented energy (Eq.~\ref{eqn : energy}) with the constraint $\mathrm{det}\mathbf{F}=1$. This last condition is ensured by adding to Eq.~\ref{eqn : energy} the supplementary term
\begin{equation}
  \int_{\Omega_0} p\left(\mathrm{det}\mathbf{F}-1\right) \mathrm{d}V_0,
\label{eqn : lagrange}
\end{equation}
where $p$ is a Lagrange multiplier to be computed together with $\mathbf{u}$. Because the solution is expected to be axially symmetric, the displacement vector is expressed in a cylindrical coordinate system as $\mathbf{u}=u_r(r,z)\mathbf{e_r}+u_z(r,z)\mathbf{e_z}$. For this two-dimensional problem in terms of $r$ and $z$, the domain $D$ we consider in the simulation is a disk of radius $R_0$ defined as $r^2+z^2<R_0^2$.

The FE formulation, implemented numerically using the \texttt{FEniCS} finite element library \cite{Fenics2012}, is here based on the research of the stationary points of the total energy functional given by Eqs. \ref{eqn : energy} with Eq.~\ref{eqn : lagrange}. The displacement vector $\mathbf{u}$ and the Lagrange multiplier $p$ are discretized using Lagrange FEs
on a triangular mesh. The nonlinear problem in the ($\mathbf{u}$, $p$) variables is solved using a Newton algorithm based on a direct parallel solver (\texttt{MUMPS}, \cite{Amestoy2001}) by setting $\mu=1$, $R_0=1$, $\Gamma=\beta$ and $\Delta \rho \omega^2 = \alpha$.

\begin{figure}[htbp]
\centering
  \includegraphics[width=0.35\textwidth]{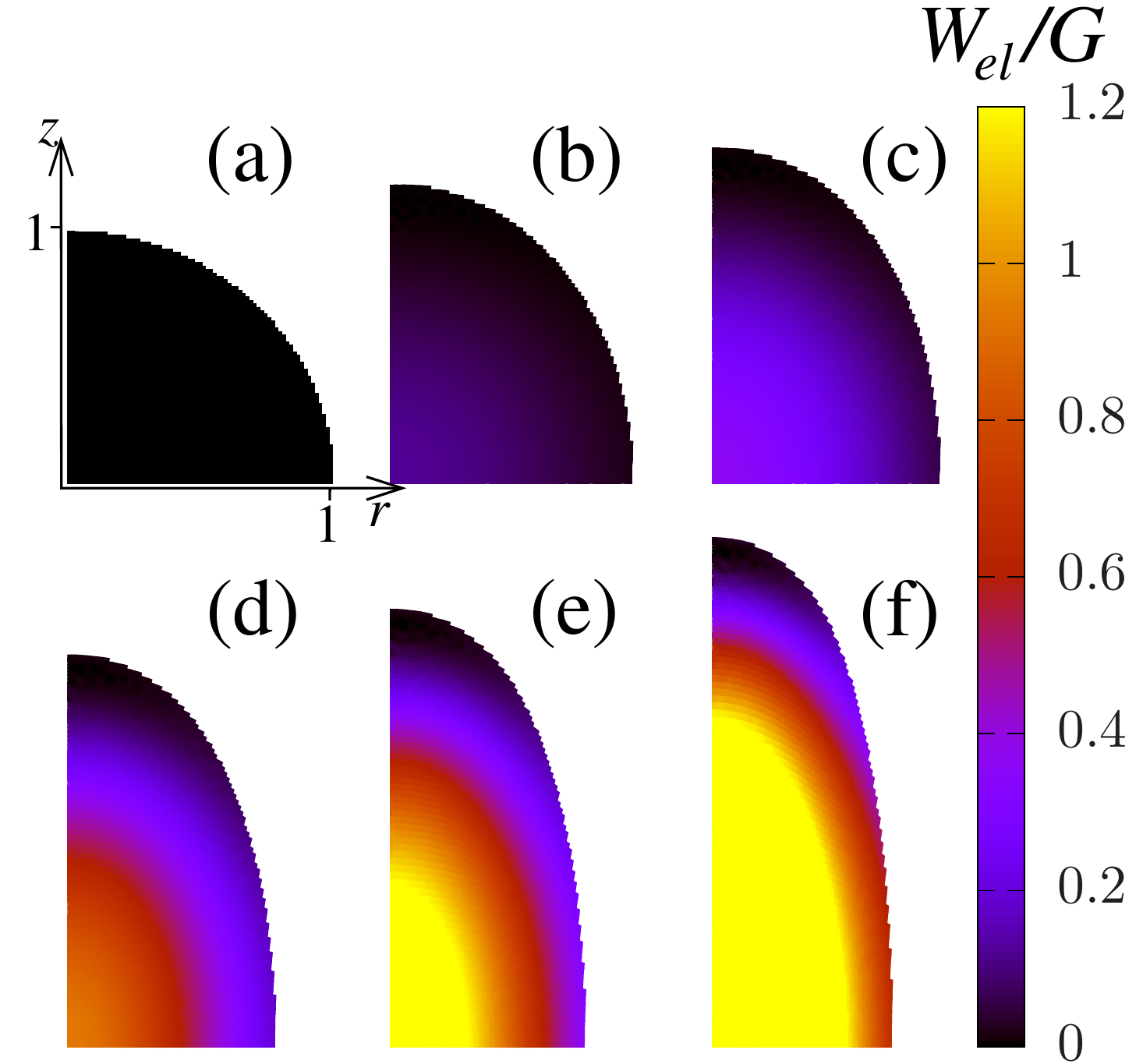}
  \caption{Maps of the reduced strain energy density $W_{el}/G_0$ computed for $\beta=1$ and different loads: $\alpha=0$ (a), $\alpha=5$ (b), $\alpha=10$ (c), $\alpha=20$ (d),$\alpha=30$ (e) and $\alpha=50$ (f). The corresponding values of $d_{max}/d_{min}$ are respectively equal to 1 (a), 1.3 (b), 1.6 (c), 2.0 (d), 2.4 (e), 3.0 (f). Unit-length is chosen so that initial configuration (a) is a disk of radius $R_0=1$. By symmetry, a quarter of the system is enough to completely characterize the deformed configurations of the bead.}
  \label{fig : map}
\end{figure}

Quasi-static simulations are computed by progressively increasing the interfacial free energy $\beta$ up to the desired dimensionless value, then by progressively incrementing the load parameter $\alpha$, recording the displacement field and the Lagrange multiplier, and reaching convergence at each step.
The equilibrium shape of the deformed body are obtained for a large range of parameters $\alpha,\beta$ (See Figs. \ref{fig : compare} and \ref{fig : map}).

Fig.~\ref{fig : map} shows that the strain energy density is inhomogeneous in the sample, whereas it was assumed to be homogeneous within the biaxial in Sec. \ref{sec : biaxial approximation}.  In Fig.~\ref{fig : compare}, the values of $d_{max}/d_{min}$ calculated from the FE method and the biaxial approximation are compared. The biaxial approximation reproduces only qualitatively the deformation behaviour for small to moderate $\beta$ (see Fig.~\ref{fig : compare} for $\beta=0$, $2$), converging quantitatively to the FE results only for larger values of $\beta$  ({\it e.g.} $\beta=4$, $6$). Indeed, the biaxial approximation considers only the average deformation in the material instead of considering the local deformation, hence the observed discrepancies. This said, even if a quantitative analysis requires the use of the more precise FE calculation, the biaxial approximation provides a rigorous basis to understand bead deformation and the role played by interfacial stresses when elastic objects get deformed.

\begin{figure}[htbp]
\centering
  \includegraphics[width=0.35\textwidth]{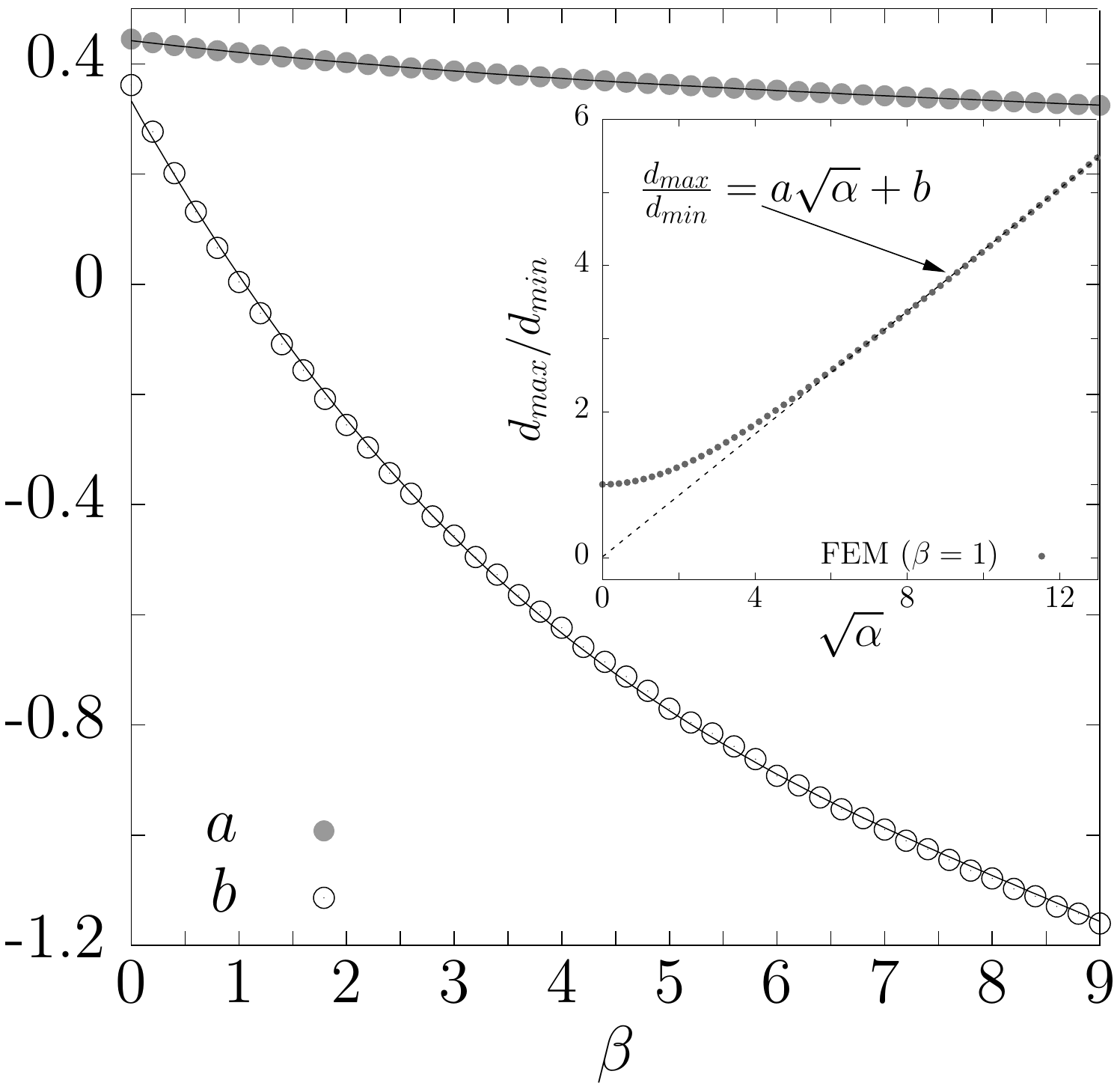}
  \caption{$a(\beta)$ and $b(\beta)$ resulting from the fits of the large deformation limit of function $a\sqrt{\alpha}+b$ on $d_{max}/d_{min}$. The range for the fits is $d_{max}/d_{min} \in [2.5,6]$, in accordance with the domain explored in experiments detailed in Sec. \ref{sec : experiments}. Solid lines result from fourth order polynomial fits for $a(\beta)$ and $b(\beta)$ (see Table \ref{tab : fits_ab}). Inset: $d_{max}/d_{min}$ calculated by the FE simulations, as a function of $\sqrt{\alpha}$ for $\beta=1$. $d_{max}/d_{min}$ is well approximated by the linear equation $d_{max}/d_{min}=a\sqrt{\alpha}+b$ in the large deformations limit.}
  \label{fig : fem}
\end{figure}

\begin{figure}[htbp]
\centering
  \includegraphics[width=0.4\textwidth]{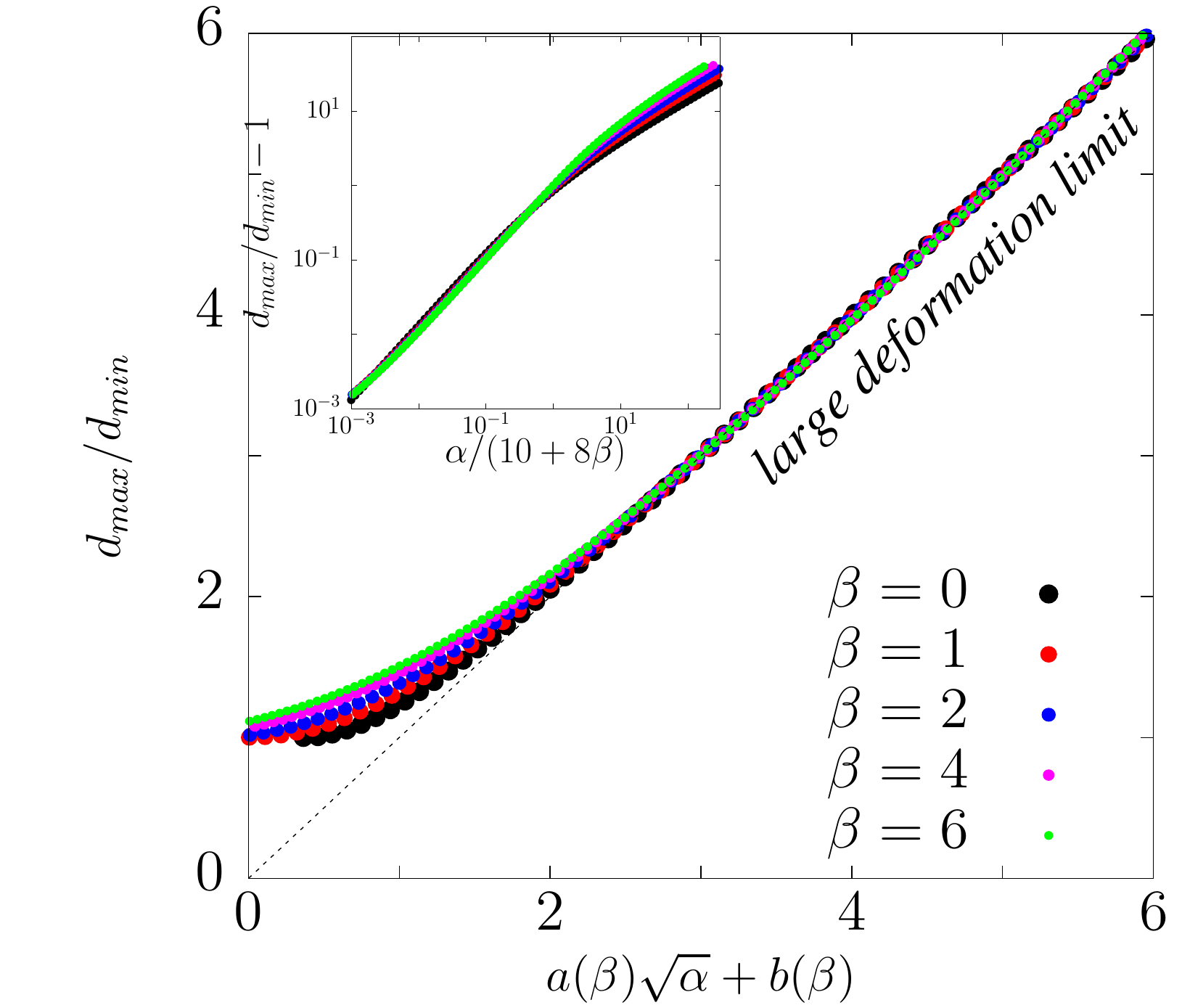}
  \caption{$d_{max}/d_{min}$ computed from the FE simulations, as a function of $a\sqrt{\alpha}+b$ where $a$ and $b$ are determined, as functions of $\beta$, in Fig.~\ref{fig : fem}. Inset: $d_{max}/d_{min}-1$ as a function of $\alpha/(10+8\beta)$.}
  \label{fig : fem collapse}
\end{figure}

Inspired by the results obtained in the framework of the biaxial approximation (Sec. \ref{sec : biaxial approximation}), we focus on the large deformation limit. Our simulations still suggest that $d_{max}/d_{min}$ behaves as
\begin{equation}
  d_{max}/d_{min}\sim  a\sqrt{\alpha}+b
  \label{eqn : asymptote}
\end{equation}
in the large deformation limit for any tested value of $\beta$ (Fig.~\ref{fig : fem}), where $a$ and $b$ are two fitting parameters. Interestingly, the variations of $b$ as a function of $\beta$ are far more pronounced than the variations of $a$, a result reminiscent with what was obtained in Sec. \ref{sec : biaxial approximation} (see Eq.~\ref{eqn:energy8}).
Let us consider now experiments in which $d_{max}/d_{min}$ has been measured as a function of $\omega$. In the regime of large deformations, we expect from Eq. \ref{eqn : asymptote} the deformed shape of the spinning bead to follow $d_{max}/d_{min} \sim A \omega + B$. This is indeed observed for our polyacrylamide millimetric particles (see Fig.~\ref{fig : exp1} and Sec. \ref{sec : experiments} for more details). Hence $A$ and $B$ can be, in principle, experimentally determined. In the other hand, we know from the results of the FE simulations that:
\begin{equation}
  d_{max}/d_{min} \sim a(\beta) \sqrt{\alpha} + b(\beta)=a(\beta)\sqrt{\frac{\Delta \rho}{G_0}}R_0\omega+b(\beta).
  \label{eqn : strategy}
\end{equation}
By identifying $A$ and $B$ within equation \ref{eqn : strategy}, we obtain:
\begin{equation}
  \left\{\begin{array}{l}
  A=a(\beta)R_0\sqrt{\frac{\Delta \rho}{G_0}}\\ B=b(\beta).
  \end{array} \right.
  \label{eqn : A B}
\end{equation}
The dependence of $b$ on $\beta$ being known from the FE solution, $\beta$ can be determined. Then, from the first equation in \ref{eqn : A B}, one can determine $G_0$ by performing a linear fit of the experimental data ($d_{max}/d_{min}$ versus $\omega$) in the large deformation limit, and finally, the interfacial free energy can be calculated as $\Gamma=\beta R_0 G_0$.
Once more, this shows that both the interfacial free energy $\Gamma$ and the shear modulus $G_0$ of the bead can be extracted by fitting the bead deformation as a function of $\omega$.
To elucidate better the validity of Eq. \ref{eqn : asymptote}, $d_{max}/d_{min}$ is plotted as a function of $a(\beta)\sqrt{\alpha}+b(\beta)$ for different values of $\beta$ (Fig.~\ref{fig : fem collapse}). $a(\beta)$ and $b(\beta)$ have been determined by considering deformations $d_{max}/d_{min}$ in the range $[2.5,6]$, accordingly with the domain explored in experiments discussed in Sec. \ref{sec : experiments}. Even if the asymptotic regime is never strictly reached in this range for any $\beta$, the linear approximation of $d_{max}/d_{min}$ as function of $\sqrt{\alpha}$ remains very good.
To make our results readily exploitable for future measurements, we have fitted separately $a(\beta)$ and $b(\beta)$ with a cubic function, namely $K_{1}+K_{2}\beta+K_{3}\beta^2+K_{4}\beta^3$. We report all values for the constant $K_{i}$ in Table \ref{tab : fits_ab} and the result of the fit is shown in Fig. \ref{fig : fem}.
\begin{table}
\begin{ruledtabular}
  \begin{tabular}{c|c|c|c|c}
 &$K_1$ &$K_2$& $K_3$& $K_4$\\\hline
$a(\beta)$ & $0.4422$ & $-0.02848$ & $\num{1.69e-3}$ & $\num{-6.65e-5}$ \\ \hline
$b(\beta)$ & $0.3327$ & $-0.3467$ & $\num{3.12e-2}$ & $\num{-1.23e-3}$ \\
\end{tabular}
  \caption{Coefficients $K_i$ giving the best cubic polynomial fit of $a(\beta)$ and $b(\beta)$ shown in Fig. \ref{fig : fem} and discussed in the main text.} \label{tab : fits_ab}
\end{ruledtabular}
\end{table}
The collapse of all curves in Fig.~\ref{fig : fem collapse} confirms that the approximation $d_{max}/d_{min} \simeq a(\beta)\sqrt{\alpha}+b(\beta)$ is relevant for $d_{max}/d_{min}$ in the experimental range $[2,6]$.

Finally it's worth noting that in the case $\alpha\ll 1$ (small deformation limit), the expression obtained from the biaxial approximation seems to hold well in the framework of the FE calculation (see inset of Fig.~\ref{fig : fem collapse}): all the deformations calculated for different $\beta$-values via FE method collapse on the bisector of the first quadrant when plotted versus the deformation obtained under biaxial approximation (Eq. \ref{eqn:energy13}). Indeed, matching the effective local deformation with the overall deformation of the bead is here relevant, because the material behaviour can be linearized within the limit of the small deformations.

\section{Experiments} \label{sec : experiments}

\subsection{Materials and Methods}

Polyacrylamide beads are prepared by copolymerization of acrylamide and N, N'−methylenebisacrylamide in the presence of Tetramethylenediamine (TEMED) and sodium persulfate as initiators, in water. Prior to mixing the constituents, all the solutions are saturated with nitrogen gas, to ensure the near insufficiency of oxygen. A given volume of the liquid mixture, corresponding to the radius of the bead, is transferred to an Eppendorf tube filled with the fluorinated oil in which all beads are solidly spun in our experiments. The aqueous droplets are small enough so that interfacial free energy ($\simeq \SI{33 \pm 3}{\milli \newton \per \meter}$ measured by SDT in absence of crosslinker) made them spherical in oil. 
The polymerization and interchain crosslinking stopped after approximately 2 hours. The crosslinker and the acrylamide monomer concentrations were fixed respectively to  $0.00119\pm0.0006$ mol/l and $0.45\pm0.01$ mol/l for all preparations. The same preparation protocol has been previously employed in our group to synthesize beads in silicon oil with shear modulus ranging from 13 Pa to 29  Pa \cite{Arora2018}. Hence hereafter we will not consider any crosslinker and/or monomer density variation in the beads, whose effects will be possibly investigated in a future publication.
Since we consider beads characterized by low mass fractions of acrylamide, their mass density can be considered equal to the density of water at $T=$ 25 $^{\circ}$C, $\rho_i=\SI{0.997}{\gram \per \cubic \centi \meter}$.

All experiments were performed with a Kr\"{u}ss spinning drop tensiometer (SDT). Rates of rotation were accurate to 1\%. The outer liquid, Fomblin Y oil [linear formula  CF$_3$O[-CF(CF$_3$)CF$_2$O-]$_x$(-CF$_2$O-)$_y$CF$_3$] of mass density $\rho_{o}=\SI{1.9}{\gram \per \cubic \centi \meter}$ was purchased from Sigma-Aldrich and used without further purification at \SI{25.0}{\celsius}. The temperature of the setup was always set to \SI{25.0\pm 0.5}{\celsius} and kept constant using a flow of temperature-controlled air.

All beads were illuminated by a blue Light Emitting Diode (LED) with a dominant emission wavelength of \SI{469}{\nano \meter}. Measurements were performed using a cylindrical capillary with internal diameter $2R_c=\SI{3.25}{\milli \meter}$.
Video recording has been performed by using a CCD camera attached to the SDT with a field of view $\SI{6}{\milli \meter}\times \SI{4.5}{\milli \meter}$ and resolution \SI{2.3}{\micro \meter}.
Different tests were performed with rotation rates ranging from 6000 rpm to 15000 rpm. For our beads/oil system the displacement of the drop off the rotation axis due to buoyancy was smaller than 7 $\mu$m for $\omega>\SI{800}{\radian \per \second}$, as calculated following Ref. \cite{currie_buoyancy_1982}. Such unavoidable deviation due to buoyancy is therefore much smaller than the bead size and of the same order of magnitude of the resolution of the camera used for the visual inspection the equilibrium shapes of the beads. The effect of buoyancy can thus be neglected and the measured deformation for $\omega>\SI{800}{\radian \per \second}$ can be considered as only originated from the balance between the external forcing and the response of the material.

Being the refractive index of the background fluid ($n_b=1.299$) close to that of water, the contrast in refractive index with the beads was not sufficient to ensure good detection of the beads boundaries. Fluorescent labelling was therefore needed to track the bead deformation. Under the illumination of the blue LED light, fluorescein-rich beads appear as bright green-yellow regions, since the fluorescein adsorption and emission spectra (in polar solvents) are peaked at $\lambda\approx \SI{485}{\nano \meter}$ and $\lambda\approx \SI{511}{\nano \meter}$ \cite{panchompoo_one-step_2012,szalay_effect_1964,carbonaro_spinning_2019}, respectively.


\begin{figure}[htbp]
\centering
  \includegraphics[width=0.4\textwidth]{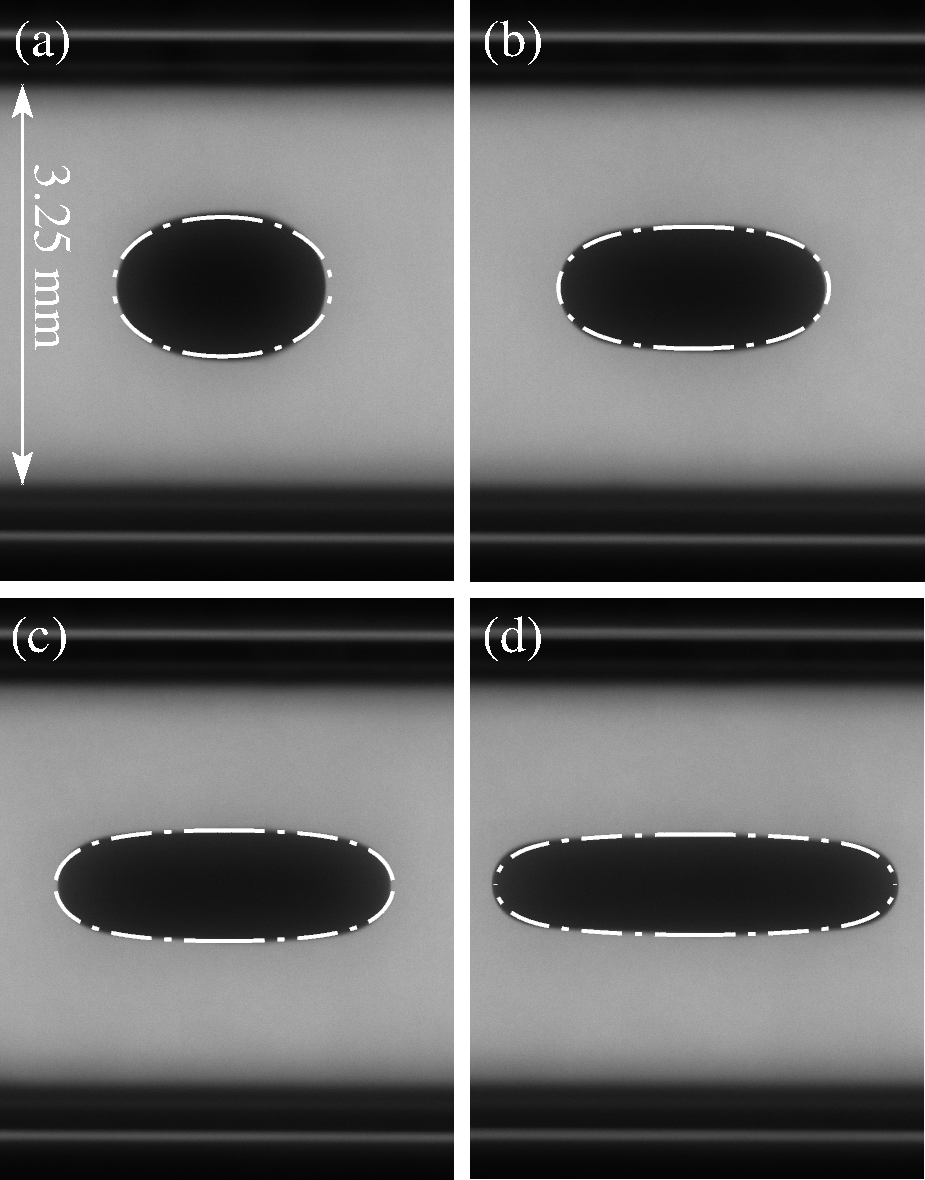}\\
  \caption{Snapshots of sample B1 (details are given in table \ref{tab : fit}) spinning with angular velocities 6000 rpm (a), 9000 rpm (b), 12000 rpm (c) and 15000 rpm (d). The observed deformation is correctly captured by the one obtained minimizing the total energy using the FE method (Sec. \ref{sec : fem}), whose result is represented by the white dash-dotted lines. The corresponding values of the load are $\alpha=27$ (a), 60 (b), 107 (c) and 167 (d). The global deformations $d_{max}/d_{min}$ are 1.6 (a), 2.2 (b), 3.0 (c) and 3.9 (d).}
  \label{fig : shape}
\end{figure}

\subsection{Analysis}
\begin{figure}[htbp]
\centering
\includegraphics[width=0.45\textwidth]{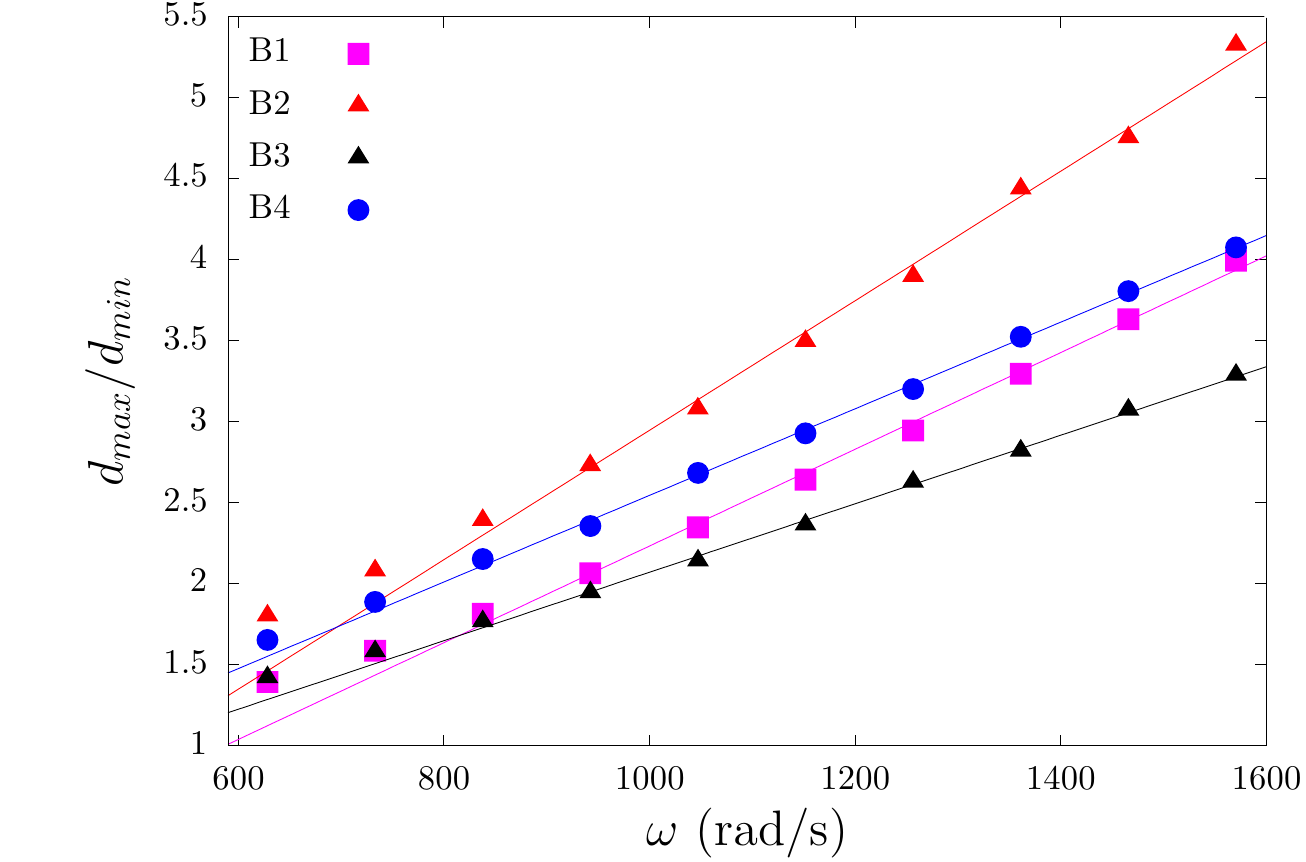}
  \caption{$d_{max}/d_{min}$ as a function of $\omega$ for four polyacrylamide beads with different radius $R_0$ and/or different shear modulus $G_0$.  Solid lines are the results of linear fit $d_{max}/d_{min}=A\omega +B$ carried out for $\omega > \SI{800}{\per \second}$ (see Table \ref{tab : fit}).  }
\label{fig : exp1}
\end{figure}

Four beads (coded as B1, B2, B3, B4) have been tested in the SDT in the large deformation limit ($d_{max}/d_{min}>2$). Figure \ref{fig : shape} shows one fluoresceinated bead (B1) under different forcing (from 6000 rpm to 15000 rpm).
For all beads 
we have extracted the parameter $A$ and $B$ from the relation $d_{max}/d_{min}=A\omega +B$ in the large deformation regime (see Fig.~\ref{fig : exp1}), and then, $G_0$ and $\Gamma$ have been deduced following the procedure detailed in Sec. \ref{sec : fem} (see Table \ref{tab : fit}).
\begin{table}
\begin{ruledtabular}
  \begin{tabular}{c|c|c|c|c|c|c|c}

 Sample   & $R_0$ &$A$       & $B$      &$\beta$  &$G_0$    &$\Gamma$ \\
          & mm &  s       &          &         & Pa      &mN/m \\\hline
 B1 &0.73&$\num{2.99e-3}$  &-0.75     &4.9      &7.1      &25.6\\
   &$\pm 0.01$&$\pm \num{0.07e-3}$ &$\pm 0.09$&$\pm 0.6$&$\pm 0.3$&$\pm 4.4$    \\\hline
 B2 &0.89&$\num{4.00e-3}$   &-1.05     &7.8     &5.1      &35.3 \\
   &$\pm 0.01$&$\pm \num{0.01e-3}$ &$\pm 0.14$&$\pm 1.6$&$\pm 0.3$&$\pm  8.9$    \\ \hline
 B3 &0.89&$\num{2.11e-3}$  &-0.042    &1.2     &27.8     &28.9 \\
   &$\pm 0.01$& $\pm \num{0.04e-3}$&$\pm 0.050$&$\pm 0.2$&$\pm 1.0$  &$\pm 5.0$\\\hline
 B4 &0.965&$\num{2.67e-3}$ &-0.13    &1.5     &16.4     &29.3 \\
   &$\pm 0.010$& $\pm \num{0.04e-3}$&$\pm 0.05$&$\pm 0.2$&$\pm 4.0$  &$\pm 10.0$
\end{tabular}
  \caption{Data obtained from the analysis of the measurements of $d_{max}/d_{min}$ as a function of $\omega$ for the four tested samples.} \label{tab : fit}
\end{ruledtabular}
\end{table}

To check further the validity of our approach for largely deformed beads, we have rescaled our experimental data with the same procedure already adopted for the theoretical values of $d_{max}/d_{min}$ (Fig. \ref{fig : fem collapse}).
\begin{figure}[htbp]
\centering
\includegraphics[width=0.45\textwidth]{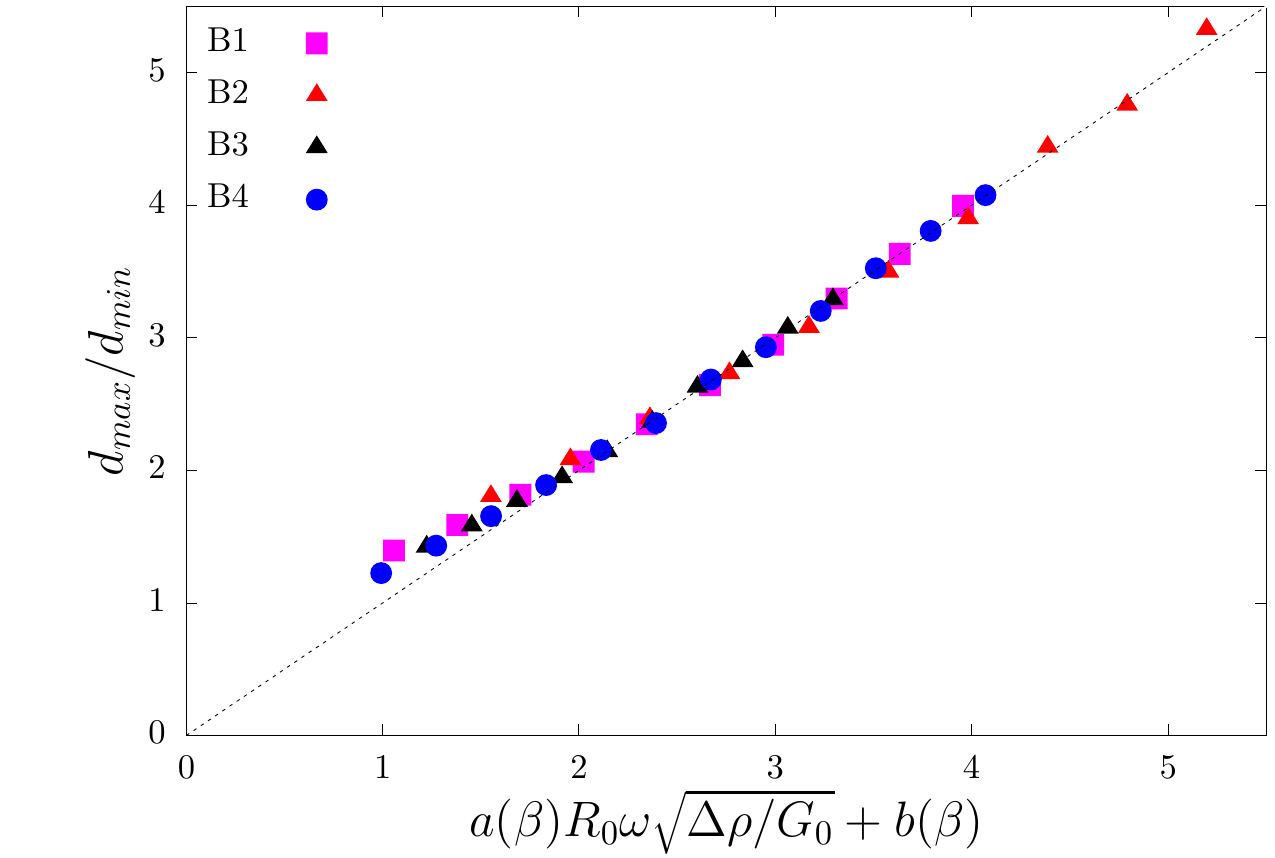}
  \caption{$d_{max}/d_{min}$ as a function of the rescaled angular velocity, for polyacrylamide beads with different radius and/or elastic modulus.}
  \label{fig : exp collapse}
\end{figure}
Figure \ref{fig : exp collapse} shows all values of $d_{max}/d_{min}$ obtained for different synthesis of polyacrylamide beads in function of the rescaled forcing $A\omega+B$. All data collapse on the same master curve,
showing that the large deformation limit is indeed reached in all cases.

The values of the interfacial free energy are found to be similar from one bead to another one, with a weighted average of $\Gamma = \SI{28.1 \pm 3.0}{\milli \newton \per \meter}$ (See Table \ref{tab : fit}). Error bars could be reduced by exploring larger values of $\omega$, which was not possible with our SDT. Interestingly, these values of $\Gamma$ are in agreement with the liquid-liquid interfacial tension $\Gamma_{liq}=\SI{33 \pm 3}{\milli \newton \per \meter}$ measured via SDT between the oil and a solution of non-crosslinked polyacrylamide polymers at the same acrylamide molar concentration of the beads. Hence, the interfacial free energy and interfacial tension of the beads are equal within our experimental uncertainties.

As a final remark we stress that the polymerization reaction used to synthesize the beads occurred into the background fluorinated oil used successively in SDT experiments. On the one hand, this ensures that the beads are not subject to a possible contamination that may arise from the synthesis in other immiscible media and that may affect successively the measurement performed in the SDT, notably the measured interfacial energy. On the other hand this oil inevitably alters the polymerization process, hence modifying the value of the gel modulus with respect to other similar synthesis already performed in our group \cite{Arora2018}. This deviation, which can be significant for gels with low elastic modulus, is more pronounced as the contact surface between the aqueous solution and oil is larger. For this reason, we could not perform different types of synthesis like those carried out to produce macroscopic polyacrylamide gels \cite{chakrabarti_elastowetting_2018, Arora2018} whose modulus can be determined via other methods \cite{Arora2018,Mora2020}. This hampered a direct cross-check of the values obtained for the shear modulus of our beads. Further research activity is being carried on in our group to develop a synthesis protocol enabling to crosscheck the measurement of elastic moduli obtained via a SDT. Despite of that, the values obtained with the SDT method seem relevant as they are in excellent agreement with those found via impact experiments \cite{Arora2018} for similar polyacrylamide beads, suggesting that the bead deformation method under centrifugal forcing may serve as an ideal strategy to measure accurately both the elastic moduli and the surface energies of soft elastic materials.

\section{Conclusions} \label{sec : conclusion}

Due to the interplay between bulk and surface forces acting simultaneously, isolating the effects of the solid-liquid interfacial free energy constant of a soft solid is challenging. While for materials with large shear moduli the contribution of interfacial stresses to deformation can generally be safely neglected, the equilibrium shapes and the stability of soft solids under external drives are altered significantly by their ability to store and/or release interfacial energy.
For such systems, measuring the shear modulus is also a difficult task since standard rheometric techniques are often confronted with experimental issues, like wall slip, edge fracture and instrumental resolution, hampering the accurate measurement of the material moduli. For this reason, a robust method able to measure unambiguously both the shear modulus and the interfacial free energy is highly desirable.
In this paper we have shown that simultaneous measurements of the shear modulus and the interfacial free energy of elastic materials can be achieved without contact with a solid surface by analysing the shape of spinning soft beads. These measurements are based on a gradual variation of the load, {\it i.e.} of the angular velocity of the bead. This method requires the prior knowledge of the constitutive equation of the material.
Here, in particular, we have investigated the case the isochoric neo-Hookean model, valid for polyacrylamide gels. We have measured the solid-liquid interfacial free energy for solid particles undergoing large deformations, and we have shown that, for these systems, the interfacial free energy is similar to the liquid-liquid interfacial tension measured in absence of elastic bulk forces. Our results corroborate a scenario where the deformation of soft amorphous polymer materials under an external load can be described considering one single interfacial free energy parameter independent on the deformation.
For materials following another known constitutive law (like the Gent model \cite{Gent1996} or Mooney-Rivlin model \cite{Mooney1940}), the method described here also applies provided that this elastic law is accounted for in the simulations so that the functions $a(\beta)$ and $b(\beta)$ are properly determined. We hope that our work motivates further research both to improve and adapt SDT apparatus to the measurement of the elastic modulus of soft materials and to generalize our results to different elastic and viscoelastic systems.

\section*{Conflicts of interest}
The authors state that there are no conflicts to declare.

\section*{Acknowledgements}
CL and C-A acknowledge  partial financial support of the  H2020 Program (Marie Curie Actions) of the European Commission's Innovative Training Networks (ITN) (H2020-MSCA-ITN-2017) under DoDyNet REA Grant Agreement (GA) N°.765811. The authors are most grateful to Jean Marc Fromental for providing technical help.

\bibliography{bead}

\begin{thebibliography}{10}

\bibitem{Style2017}
R.~Style, A.~Jagota, C.~Hui, and E.~Dufresne, ``Elastocapillarity: surface
  tension and the mechanics of soft solids,'' {\em Annu. Rev. Condens. Matter
  Phys.}, vol.~8, pp.~99--118, 2017.

\bibitem{Bico2018}
J.~Bico, E.~Reyssat, and B.~Roman, ``Elastocapillarity: When surface tension
  deforms elastic solids,'' {\em Annual Review of Fluid Mechanics}, vol.~50,
  pp.~629--659, 2018.

\bibitem{creton_fracture_2016}
C.~Creton and M.~Ciccotti, ``Fracture and adhesion of soft materials: a
  review,'' {\em Rep. Prog. Phys.}, vol.~79, no.~4, p.~046601, 2016.

\bibitem{Nicolson1955}
M.~Nicolson, ``Surface tension in ionic crystals,'' {\em P. Roy. Soc. A-Math.
  Phy.}, vol.~228, pp.~490--510, 1955.

\bibitem{Mora_prl2010}
S.~Mora, T.~Phou, J.~M. Fromental, L.~M. Pismen, and Y.~Pomeau, ``Capillarity
  driven instability of a soft solid,'' {\em Phys. Rev. Lett.}, vol.~105,
  p.~214301, 2010.

\bibitem{Style2013}
R.~Style, C.~Hyland, R.~Boltyanskiy, J.~Wettlaufer, and E.~Dufresne, ``Surface
  tension and contact with soft elastic solids,'' {\em Nat. Commun.}, vol.~4,
  p.~2728, 2013.

\bibitem{andreotti_elastocapillary_2011}
B.~Andreotti, A.~Marchand, S.~Das, and J.~H. Snoeijer, ``Elastocapillary
  instability under partial wetting conditions: {Bending} versus buckling,''
  {\em Phys. Rev. E}, vol.~84, no.~6, 2011.

\bibitem{evans_elastocapillary_2013}
A.~A. Evans, S.~E. Spagnolie, D.~Bartolo, and E.~Lauga, ``Elastocapillary
  self-folding: buckling, wrinkling, and collapse of floating filaments,'' {\em
  Soft Matter}, vol.~9, no.~5, pp.~1711--1720, 2013.

\bibitem{chakrabarti_direct_2013}
A.~Chakrabarti and M.~K. Chaudhury, ``Direct {Measurement} of the {Surface}
  {Tension} of a {Soft} {Elastic} {Hydrogel}: {Exploration} of
  {Elastocapillary} {Instability} in {Adhesion},'' {\em Langmuir}, vol.~29,
  no.~23, pp.~6926--6935, 2013.

\bibitem{Arora2018}
S.~Arora, J.~Fromental, S.~Mora, T.~Phou, and C.~Ligoure, ``Impact of beads and
  drops on a repellent solid surface: a unified description,'' {\em Phys. Rev.
  Lett.}, vol.~120, p.~148003, 2018.

\bibitem{Limat2018}
M.~Zhao, F.~Lequeux, T.~Narita, M.~Roch\'e, and L.~Limat, ``Growth and
  relaxation of a ridge on a soft poroelastic substrate,'' {\em Soft Matter},
  vol.~14, pp.~61--72, 2018.

\bibitem{mondal_estimation_2015}
S.~Mondal, M.~Phukan, and A.~Ghatak, ``Estimation of solid-liquid interfacial
  tension using curved surface of a soft solid,'' {\em Proc. Natl. Acad. Sci.
  U.S.A.}, vol.~112, no.~41, pp.~12563--12568, 2015.

\bibitem{Andreotti2016}
B.~Andreotti and J.~Snoeijer, ``Soft wetting and the shuttleworth effect, at
  the crossroads between thermodynamics and mechanics,'' {\em EPL-EuroPhys.
  Lett.}, vol.~113, p.~66001, 2016.

\bibitem{Shuttleworth1950}
R.~Shuttleworth, ``The surface tension of solids.,'' {\em Proc. Phys. Soc.},
  vol.~63, pp.~444--457, 1950.

\bibitem{muller_elastic_2004}
P.~Muller, ``Elastic effects on surface physics,'' {\em Surf. Sci. Rep.},
  vol.~54, no.~5-8, pp.~157--258, 2004.

\bibitem{orowan_surface_1970}
E.~Orowan, ``Surface energy and surface tension in solids and liquids,'' {\em
  P. Roy. Soc A-Math. Phys}, vol.~316, no.~1527, pp.~473--491, 1970.

\bibitem{savina_faceting_2003}
T.~V. Savina, A.~A. Golovin, S.~H. Davis, A.~A. Nepomnyashchy, and P.~W.
  Voorhees, ``Faceting of a growing crystal surface by surface diffusion,''
  {\em Phys. Rev. E}, vol.~67, no.~2, 2003.

\bibitem{vaidya_synthesis_2002}
A.~Vaidya and M.~K. Chaudhury, ``Synthesis and {Surface} {Properties} of
  {Environmentally} {Responsive} {Segmented} {Polyurethanes},'' {\em J. Colloid
  Interface Sci.}, vol.~249, no.~1, pp.~235--245, 2002.

\bibitem{hillborg_nanoscale_2004}
H.~Hillborg, N.~Tomczak, A.~Olàh, H.~Schönherr, and G.~J. Vancso, ``Nanoscale
  {Hydrophobic} {Recovery}: {A} {Chemical} {Force} {Microscopy} {Study} of
  {UV}/{Ozone}-{Treated} {Cross}-{Linked} {Poly}(dimethylsiloxane),'' {\em
  Langmuir}, vol.~20, no.~3, pp.~785--794, 2004.

\bibitem{style_elastocapillarity_2017}
R.~W. Style, A.~Jagota, C.-Y. Hui, and E.~R. Dufresne, ``Elastocapillarity:
  {Surface} {Tension} and the {Mechanics} of {Soft} {Solids},'' {\em Annu. Rev.
  Condens. Matter Phys.}, vol.~8, no.~1, pp.~99--118, 2017.

\bibitem{andreotti_statics_2020}
B.~Andreotti and J.~H. Snoeijer, ``Statics and {Dynamics} of {Soft}
  {Wetting},'' {\em Annu. Rev. Fluid Mech.}, vol.~52, no.~1, pp.~285--308,
  2020.

\bibitem{Mora_softmatter2011}
S.~Mora, M.~Abkarian, H.~Tabuteau, and Y.~Pomeau, ``Surface instability of soft
  solids under strain,'' {\em Soft Matter}, vol.~7, pp.~10612--10619, 2011.

\bibitem{Hui2002}
C.~Hui, A.~Jogota, Y.~Lin, and E.~Kramer, ``Constraints on microcontact
  printing imposed by stamp deformation,'' {\em Langmuir}, vol.~18,
  pp.~1394--1407, 2002.

\bibitem{Mora_prl2013}
S.~Mora, C.~Maurini, T.~Phou, J.~M. Fromental, B.~Audoly, and Y.~Pomeau,
  ``Solid drops: Large capillary deformations of immersed elastic rods.,'' {\em
  Phys. Rev. Lett.}, vol.~111, p.~114301, 2013.

\bibitem{Mora_JPhys2015}
S.~Mora and Y.~Pomeau, ``Softening of edges of solids by surface tension,''
  {\em J. Phys. Condens. Matter}, vol.~27, p.~194112, 2015.

\bibitem{Jagota2014}
D.~Paretkar, X.~Xu, C.~Y. Hui, and A.~Jagota, ``Flattening of a patterned
  compliant solid by surface stress,'' {\em Soft Matter}, vol.~10,
  pp.~4084--4090, 2014.

\bibitem{Chakrabarti2016}
A.~Chakrabarti, M.~K. Chaudhury, S.~Mora, and Y.~Pomeau, ``Elastobuoyant heavy
  spheres: A unique way to study nonlinear elasticity,'' {\em Phys. Rev. X},
  vol.~6, p.~041066, 2016.

\bibitem{Delavoipiere2016}
J.~Delavoipiere, Y.~Tran, E.~Verneuil, and A.~Chateauminois, ``Poroelastic
  indentation of mechanically confined hydrogel layers,'' {\em Soft Matter},
  vol.~12, p.~8049, 2016.

\bibitem{vonnegut_rotating_1942}
B.~Vonnegut, ``Rotating {Bubble} {Method} for the {Determination} of {Surface}
  and {Interfacial} {Tensions},'' {\em Rev. Sci. Instrum.}, vol.~13, no.~1,
  pp.~6--9, 1942.

\bibitem{bamberger_effects_1984}
S.~Bamberger, G.~V. Seaman, K.~Sharp, and D.~E. Brooks, ``The effects of salts
  on the interfacial tension of aqueous dextran poly(ethylene glycol) phase
  systems,'' {\em J. Colloid Interface Sci.}, vol.~99, no.~1, pp.~194--200,
  1984.

\bibitem{liu_concentration_2012}
Y.~Liu, R.~Lipowsky, and R.~Dimova, ``Concentration {Dependence} of the
  {Interfacial} {Tension} for {Aqueous} {Two}-{Phase} {Polymer} {Solutions} of
  {Dextran} and {Polyethylene} {Glycol},'' {\em Langmuir}, vol.~28, no.~8,
  pp.~3831--3839, 2012.

\bibitem{carbonaro_spinning_2019}
A.~Carbonaro, L.~Cipelletti, and D.~Truzzolillo, ``Spinning {Drop} {Dynamics}
  in {Miscible} and {Immiscible} {Environments},'' {\em Langmuir}, vol.~35,
  no.~35, pp.~11330--11339, 2019.

\bibitem{zoltowski_evidence_2007}
B.~Zoltowski, Y.~Chekanov, J.~Masere, J.~A. Pojman, and V.~Volpert, ``Evidence
  for the {Existence} of an {Effective} {Interfacial} {Tension} between
  {Miscible} {Fluids}. 2. {Dodecyl} {Acrylate}−{Poly}(dodecyl acrylate) in a
  {Spinning} {Drop} {Tensiometer},'' {\em Langmuir}, vol.~23, no.~10,
  pp.~5522--5531, 2007.

\bibitem{pojman_evidence_2006}
J.~A. Pojman, C.~Whitmore, M.~L. Turco~Liveri, R.~Lombardo, J.~Marszalek,
  R.~Parker, and B.~Zoltowski, ``Evidence for the {Existence} of an {Effective}
  {Interfacial} {Tension} between {Miscible} {Fluids}: {Isobutyric}
  {Acid}−{Water} and 1-{Butanol}−{Water} in a {Spinning}-{Drop}
  {Tensiometer},'' {\em Langmuir}, vol.~22, no.~6, pp.~2569--2577, 2006.

\bibitem{pieper_deformation_1998}
G.~Pieper, H.~Rehage, and D.~Barthès-Biesel, ``Deformation of a {Capsule} in a
  {Spinning} {Drop} {Apparatus},'' {\em J. Colloid Interface Sci.}, vol.~202,
  no.~2, pp.~293--300, 1998.

\bibitem{joseph_spinning_1992}
D.~D. Joseph, M.~S. Arney, G.~Gillberg, H.~Hu, D.~Hultman, C.~Verdier, and
  T.~M. Vinagre, ``A spinning drop tensioextensometer,'' {\em J. Rheol.},
  vol.~36, no.~4, pp.~621--662, 1992.

\bibitem{patterson_measurement_2007}
H.~T. Patterson, K.~H. Hu, and T.~H. Grindstaff, ``Measurement of interfacial
  and surface tensions in polymer systems,'' {\em J. Polym. Sci., Polym.
  symp.}, vol.~34, no.~1, pp.~31--43, 2007.

\bibitem{Ogden1984}
R.~Ogden, {\em Non-Linear Elastic Deformations}.
\newblock Chichester: Ellis Horwood Limited, 1984.

\bibitem{Suo2012}
J.~Li, Y.~Hu, J.~Vlassak, and Z.~Suo, ``Experimental determination of equations
  of state for ideal elastomeric gels,'' {\em Soft Matter}, vol.~8, p.~8121,
  2012.

\bibitem{Mora2020}
S.~Mora, E.~Ando, J.~Fromental, T.~Phou, and Y.~Pomeau, ``The shape of hanging
  cylinders,'' {\em Soft Matter}, vol.~15, pp.~5464--5473, 2020.

\bibitem{Richard2018}
F.~Richard, A.~Chakrabarti, B.~Audoly, Y.~Pomeau, and S.~Mora, ``Buckling of a
  spinning elastic cylinder: linear, weakly nonlinear and post-buckling
  analyses,'' {\em Proc. R. Soc. A}, vol.~474, p.~20180242, 2018.

\bibitem{Fenics2012}
A.~Logg, K.~Mardal, and G.~Wells, {\em Automated Solution of Differential
  Equations by the Finite Element Method}.
\newblock Springer, 2012.

\bibitem{Amestoy2001}
P.~Amestoy, I.~Duff, J.~L'Escellent, and J.~Koster, ``A fully asynchronous
  multifrontal solver using distributed dynamic scheduling,'' {\em SIAM J.
  Matrix Anal. Appl.}, vol.~23, pp.~15--41, 2001.

\bibitem{currie_buoyancy_1982}
P.~Currie and J.~Van~Nieuwkoop, ``Buoyancy effects in the spinning-drop
  interfacial tensiometer,'' {\em J. Colloid Interface Sci.}, vol.~87, no.~2,
  pp.~301--316, 1982.

\bibitem{panchompoo_one-step_2012}
J.~Panchompoo, L.~Aldous, M.~Baker, M.~I. Wallace, and R.~G. Compton,
  ``One-step synthesis of fluorescein modified nano-carbon for {Pd}(ii)
  detection via fluorescence quenching,'' {\em Analyst}, vol.~137, no.~9,
  p.~2054, 2012.

\bibitem{szalay_effect_1964}
L.~Szalay and E.~Tombácz, ``Effect of the solvent on the fluorescence spectrum
  of trypaflavine and fluorescein,'' {\em Acta Phys. Acad. Sci. Hung.},
  vol.~16, no.~4, pp.~367--371, 1964.

\bibitem{chakrabarti_elastowetting_2018}
A.~Chakrabarti, A.~Porat, E.~Raphaël, T.~Salez, and M.~K. Chaudhury,
  ``Elastowetting of {Soft} {Hydrogel} {Spheres},'' {\em Langmuir}, vol.~34,
  no.~13, pp.~3894--3900, 2018.

\bibitem{Gent1996}
A.~Gent, ``A new constitutive relation for rubber.,'' {\em Rub. Chem. Tech.},
  vol.~69, pp.~59--61, 1996.

\bibitem{Mooney1940}
M.~Mooney, ``A theory of large elastic deformation,'' {\em J. Appl. Phys.},
  vol.~11, pp.~582--592, 1940.

\end{thebibliography}
\end{document}